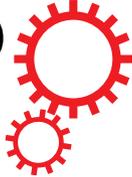

# OPEN  Origin of interfacial perpendicular magnetic anisotropy in MgO/CoFe/metallic capping layer structures

Shouzhong Peng[1,2], Mengxing Wang[1,2], Hongxin Yang[3], Lang Zeng[1,2], Jiang Nan[1,2], Jiaqi Zhou[1,2], Youguang Zhang[1,2], Ali Hallal[3], Mairbek Chshiev[3], Kang L. Wang[4], Qianfan Zhang[5] & Weisheng Zhao[1,2]

Spin-transfer-torque magnetic random access memory (STT-MRAM) attracts extensive attentions due to its non-volatility, high density and low power consumption. The core device in STT-MRAM is CoFeB/MgO-based magnetic tunnel junction (MTJ), which possesses a high tunnel magnetoresistance ratio as well as a large value of perpendicular magnetic anisotropy (PMA). It has been experimentally proven that a capping layer coating on CoFeB layer is essential to obtain a strong PMA. However, the physical mechanism of such effect remains unclear. In this paper, we investigate the origin of the PMA in MgO/CoFe/metallic capping layer structures by using a first-principles computation scheme. The trend of PMA variation with different capping materials agrees well with experimental results. We find that interfacial PMA in the three-layer structures comes from both the MgO/CoFe and CoFe/capping layer interfaces, which can be analyzed separately. Furthermore, the PMAs in the CoFe/capping layer interfaces are analyzed through resolving the magnetic anisotropy energy by layer and orbital. The variation of PMA with different capping materials is attributed to the different hybridizations of both *d* and *p* orbitals via spin-orbit coupling. This work can significantly benefit the research and development of nanoscale STT-MRAM.

There is currently intense interest in magnetic tunnel junction (MTJ) with perpendicular magnetic anisotropy (PMA) for its potential to build low-power-consumption and high-density spin-transfer-torque magnetic random access memory (STT-MRAM)[1–7]. A strong PMA is required to obtain a high enough thermal stability so that data in STT-MRAM can be stored for 10 years. A milestone in this field is the discovery of interfacial PMA in MgO/CoFeB/Ta-based MTJ, which exhibited a high tunnel magnetoresistance (TMR) ratio of 120% and a low threshold switching current[8]. The origin of the interfacial PMA in the CoFe/MgO interface has been widely discussed and partially attributed to the interfacial symmetry break and Fe(Co) 3$d$- O 2$p$ orbitals hybridization[9–11].

Further experiments revealed that a capping or seed layer adjacent to CoFeB has an essential influence on the PMA value, e.g. by replacing Ta with Hf as a capping or seed layer, the interfacial PMA increases from 1.8 erg/cm$^2$ to 2.3 erg/cm$^{2,12,13}$, whereas it dramatically decreases using Ru film[14]. However, the physical mechanism behind this phenomenon remains unclear[15,16]. A series of papers have theoretically simulated and discussed magnetic anisotropy in Fe/non-magnetic metal structures and provided some interesting results. An extremely large PMA is predicted at the Fe/Ir system, while no study was performed with the most commonly used capping material Ta[17]. Magnetic anisotropies at interfaces of Fe and various non-magnetic metal elements are investigated with first-principles theory[18], whereas some results did not agree with experimental results[8,19]. Besides, the effect of MgO was rarely included in these works, which makes it difficult to compare calculation results with experimental results. Therefore, it is urgently demanded to clarify the operation principles for the PMA enhancement to fulfill the requirements of STT-MRAM applications.

In this paper, we use density functional theory (DFT) to calculate the interfacial magnetic anisotropy energy (MAE) of MgO/CoFe/X (X = Ru, Ta and Hf) structures. The trend of MAE variation with different capping materials agrees with experimental results. In order to figure out the exact origin of interfacial PMA, we analyze both the MgO/CoFe and CoFe/X interfaces with projected density of states (PDOS). Moreover, physical mechanism of PMA at the CoFe/X system is investigated with layer- and orbital-resolved MAE. Based on the results, the changes of PMA with different capping materials are mainly attributed to the MAE variations of the interfacial Co and X atoms. The $p$ orbitals of the interfacial Ta and Hf atoms are found to make significant contributions to the anisotropy.

[1]Fert Beijing Institute, Beihang University, Beijing 100191, China. [2]School of Electronic and Information Engineering, Beihang University, Beijing 100191, China. [3]Univ. Grenoble Alpes, INAC-SPINTEC, F-38000 Grenoble, France; CEA, INAC-SPINTEC, F-38000 Grenoble, France and CNRS, SPINTEC, F-38000 Grenoble, France. [4]Department of Electrical Engineering, University of California, Los Angeles, California 90095, USA. [5]School of Materials Science and Engineering, Beihang University, Beijing 100191, China. Correspondence and requests for materials should be addressed to W.Z. (email: weisheng.zhao@buaa.edu.cn) or Q.Z. (email: qianfan@buaa.edu.cn)

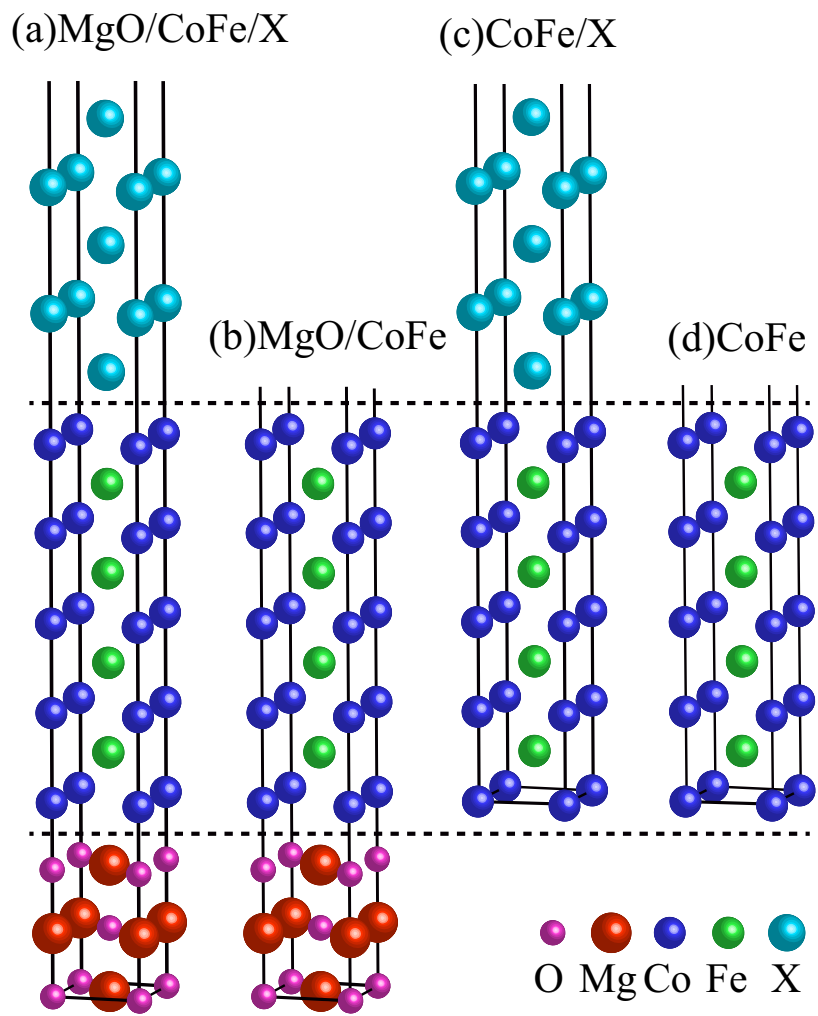

**Figure 1. Schematics of crystalline structures for (a) MgO/CoFe/X, (b) MgO/CoFe, (c) CoFe/X and (d) CoFe thin film.** A 15 Å vacuum layer is included on top of all the structures.

### Results and Discussion

Figure 1 illustrates the crystalline structures of MgO/CoFe/X, MgO/CoFe, CoFe/X and CoFe thin film, where three MgO monolayers, nine CoFe monolayers and five X monolayers are included with X denoting different capping layer atoms. The three MgO monolayers are confirmed to be sufficient in both experiments[20] and calculations[21], while the thickness of nine CoFe monolayers is about 1.3 nm, which accords well with experiments[8]. Then, we calculate the interfacial magnetic anisotropy constant $K_i$ of the MgO/CoFe/X structures with X including Ru, Ta and Hf, which are the capping materials most commonly used in experiments. The calculation results are presented in the last column of Table 1. We can find that different capping materials lead to very different PMA values. To be specific, the Ta capping layer induces a stronger PMA than the Ru, which agrees with the experimental result reported in Ref. 19. Also, it should be noted that the MgO/CoFe/X structure with X being the capping layer and the X/CoFe/MgO structure with X being the seed layer are the same in our calculations. So the phenomenon that Hf layer has a stronger enhancement effect on PMA than Ta in our calculations is coincident with experimental result reported in Ref. 12, where the Ta and Hf are used as the seed layer. Among these three systems, MgO/CoFe/Hf structure shows the largest PMA value (2.28 erg/cm$^2$), which makes Hf a potential candidate for the capping or seed layer in perpendicular MTJs with high thermal stability.

In order to investigate the origin of PMA in the MgO/CoFe/X structure, we study the magnetic anisotropy at the interfaces. As there are two ferromagnetic film/non-ferromagnetic film interfaces in this structure which may induce magnetic anisotropy, we calculate the MAEs in the MgO/CoFe and CoFe/X structures and then extract the interfacial MAEs by subtracting the MAE on the CoFe surface, which is obtained as half of the MAE in CoFe thin film due to the existence of two surfaces in this structure. The MAEs at these two interfaces are then added up to compare with the overall MAE of the three-layer system. In Table 1, we present all the results for three kinds of capping materials. It can be found that the sum of PMAs in these two interfaces approximately equals to the PMA in the MgO/CoFe/X system, which verifies that interfacial PMA in the MgO/CoFe/X system comes from these two interfaces. Moreover, it indicates that the coupling between these two interfaces is negligible, which will be verified again in the following PDOS analysis. Thus, a higher PMA in the MgO/CoFe/X system can be expected by using a proper capping material X with a strong PMA at the CoFe/X interface.

| X | CoFe surface | MgO/CoFe interface | CoFe/X interface | Sum of MAE in two interfaces | MgO/CoFe/X structure |
|---|---|---|---|---|---|
| Ru | 0.41 | 0.57 | 0.52 | 1.09 | 0.98 |
| Ta | 0.41 | 0.57 | 1.13 | 1.70 | 1.77 |
| Hf | 0.41 | 0.57 | 1.65 | 2.22 | 2.28 |

**Table 1. Calculated MAE values (erg/cm$^2$) for different structures.** The second last column shows the sum from the MAE at the MgO/CoFe interface and CoFe/X interface, where X includes Ru, Ta and Hf.

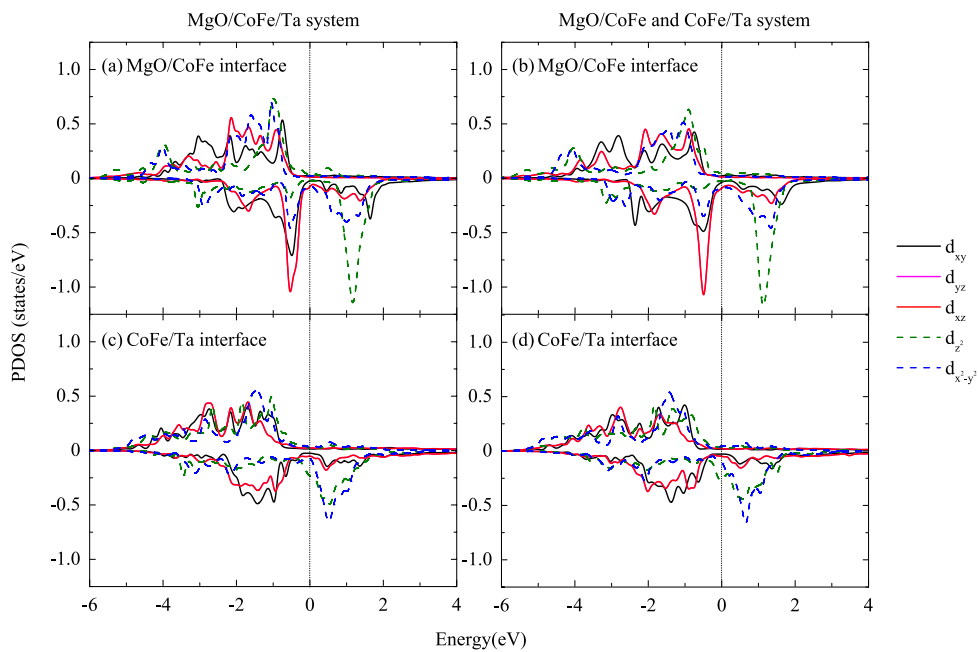

**Figure 2. Majority-spin (positive) and minority-spin (negative) PDOS on the *d* orbitals of Co atom in the MgO/CoFe interface of (a) MgO/CoFe/Ta system and (b) MgO/CoFe system, and in the CoFe/Ta interface of (c) MgO/CoFe/Ta system and (d) CoFe/Ta system.** The zero of energy is set to be $E_F$.

The PDOS analysis is performed to further explore the interaction between the MgO/CoFe and CoFe/X interfaces. In Fig. 2(a,b), we present the PDOS on the *d* orbitals of Co atoms in the MgO/CoFe interface with and without a Ta capping layer. The $d_{yz}$ and $d_{xz}$ orbitals are degenerate by the structural symmetry and their PDOSs are identical. By comparing these two figures, we can find that, in the vicinity of the Fermi energy ($E_F$), the PDOS on the *d* orbitals of Co atoms in the MgO/CoFe interface of the MgO/CoFe/Ta structure is almost the same with that in the MgO/CoFe structure, which clearly proves that the Ta capping layer has little influence on the CoFe/MgO interface in the three-layer structure. A similar conclusion that the MgO layer has almost no influence on the CoFe/Ta interface can be drawn by comparing Fig. 2(c,d). Moreover, similar results can be observed when the Ta is replaced by Ru and Hf. These phenomena are easy to understand. When the CoFe layer is thin (for example, three monolayers), there is coupling between the MgO and capping layer (See Supplementary Information, Fig. S1). However, a thick enough CoFe layer (nine monolayers, about 1.3 nm in our calculations) will block the coupling between the MgO and capping layer, so that the MgO/CoFe interface is independent of the CoFe/X interface. As a consequence, it is feasible to separate the two interfaces in the three-layer structure and analyze them separately. Since it has been proven that the PMA in Co(Fe)/MgO interface originates from the overlap between the interfacial O-$p_z$ orbital and the Co(Fe) 3*d* orbitals[9,11], here we only investigate the physical mechanism of PMA in the CoFe/X structures and compare different PMAs when the capping layer X changes.

In order to quantitatively investigate the origin of PMA, the layer- and orbital-resolved MAE is investigated with spin-orbit coupling (SOC) considered in the calculations[21,22]. In Fig. 3 we present the onsite projected MAE of the CoFe/X systems with different capping materials. Large contributions to MAE can be found at the CoFe surface (layer 1) and the CoFe/X interface (layer 9 and 10). In addition, the contributions from the CoFe surface in these three systems are almost the same, which further confirms that the capping layers have no effect on the CoFe surface. Regarding to the CoFe/X interface, different capping materials lead to much different MAE values. For the CoFe/Ta and CoFe/Hf systems, positive values of about 0.23 erg/cm$^2$ and more than 0.55 erg/cm$^2$ can be found at the interfacial Co and X atoms, respectively. However, in the CoFe/Ru system, the interfacial Co atoms have little contribution to PMA, while the interfacial Ru atoms induce an in-plane anisotropy ($-0.20$ erg/cm$^2$). These differences at the interfaces result in a much lower PMA value for the Ru-capped system. Though a stronger PMA at the interfacial Ta atoms can be found than that of the interfacial Hf atoms, the second Ta layer from the interface contributes to a strong in-plane anisotropy ($-0.57$ erg/cm$^2$), resulting in a lower PMA value than the CoFe/Hf system.

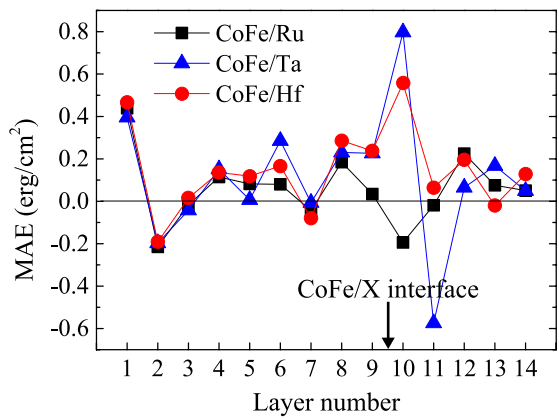

**Figure 3. Layer-resolved MAE of CoFe/X systems with different capping materials.** Nine CoFe monolayers, five X monolayers and a vacuum layer are included in the structures (as shown in Fig. 1(c)).

Then, a detailed orbital-resolved analysis is performed for the interfacial Co and X atoms as MAEs at these two layers vary greatly. In Fig. 4 we show the orbital-resolved MAE of the interfacial Co atoms, where matrix elements, for example $(d_{xy}, d_{x^2-y^2})$, denote the hybridization between two orbitals via SOC[23]. One can see clearly that the largest positive contribution to PMA for the Ta- and Hf-capped system comes from the matrix element $(d_{xy}, d_{x^2-y^2})$. However, this matrix element diminishes a lot for the Ru-capped system. Another obvious change is the decrease of the matrix element $(d_{yz}, d_{z^2})$ in the CoFe/Ru systems than that of the CoFe/Ta and CoFe/Hf systems. These changes lead to a higher PMA at the interfacial Co atoms in the Ta- and Hf-capped systems than that of the Ru-capped system in spite of a larger value of the matrix element $(d_{yz}, d_{xz})$ for the Ru-capped system.

The orbital-resolved MAE of the interfacial X atoms is shown in Fig. 5. For the interfacial Ru atoms, the primary contribution to MAE comes from the $d$ orbitals, which lead to an in-plane anisotropy due to the negative values of the matrix elements $(d_{xy}, d_{x^2-y^2})$ and $(d_{xy}, d_{xz})$. However, the situations are quite different for CoFe/Ta and CoFe/Hf systems. For interfacial X atoms in these two systems, the contributions from the $d$ orbitals are relatively small (0.15 erg/cm$^2$ and 0.02 erg/cm$^2$ for interfacial Ta and Hf, respectively), while the $p$ orbitals make significant contributions to PMA (0.64 erg/cm$^2$ and 0.54 erg/cm$^2$ for interfacial Ta and Hf, respectively). In Fig. 5(b,c) we can see that strong PMAs mainly arise from the matrix element $(p_y, p_z)$, which is enhanced by the strong SOC of the Ta and Hf atoms.

In conclusion, the interfacial PMA in MgO/CoFe/X (X = Ru, Ta and Hf) structures is calculated with first-principles theory. We confirm that PMA in these three-layer structures can be divided into two parts and analyzed separately, which indicates a simpler way to find a better material for the capping layer. Moreover, the origin of PMA in the CoFe/X interface is investigated through evaluation of layer- and orbital-resolved MAE. The changes of MAEs with different capping materials are mainly attributed to the variations of the matrix elements at the interfacial Co and X atoms. This work shows the possibility to tune PMA for different applications by choosing a proper capping material. Also, it can benefit the design of PMA-based MTJs with high thermal stability for advanced node STT-MRAM.

## Methods

First-principles calculations in this paper are based on the Vienna *ab initio* simulation package (VASP)[24–26]. A plane wave basis set and projector augmented wave (PAW) potentials are utilized. In systems shown in Fig. 1, CoFe has a CsCl structure with the CoFe [100] parallel to the MgO [110] direction. Co atoms sit atop the O atoms at the interface as reported in Ref. 11 and 27. We employ a face centre cubic (fcc) structure for the capping layer atoms X with X (100) deposited on CoFe (100) surface so as to minimize the mismatch with the CoFe layer[18,28]. For the CoFe/X interface, a Co-terminated hollow structure is used as it is the most energetically favourable structure (See Supplementary Information). All these structures are fully relaxed until the residual forces are less than 0.01 eV/Å. In all calculations, we use a cut-off energy of 520 eV and a K-point mesh of $20 \times 20 \times 1$, which is sufficient to ensure a good convergence of the MAE. Moreover, in order to prevent errors induced by dipole moment, we apply dipole corrections along the longitudinal axis[29]. Eventually, MAE is obtained by taking the energy difference when the magnetization orients along the in-plane [100] and out-of-plane [001] direction with SOC included in our calculations, where the spin-orbit term is evaluated using the second-order approximation[30] implemented in VASP:

$$H_{SOC} = \frac{\hbar^2}{4m^2c^2} \frac{1}{r} \frac{\partial V}{\partial r} \vec{L} \cdot \vec{s} \tag{1}$$

where $V$ denotes the spherical part of the all-electron Kohn-Sham potential inside the PAW spheres, while $\vec{L}$ and $\vec{s}$ represent the angular-momentum operator and the Pauli spin matrices, respectively. The layer- and orbital-resolved MAE can then be extracted from the VASP results.

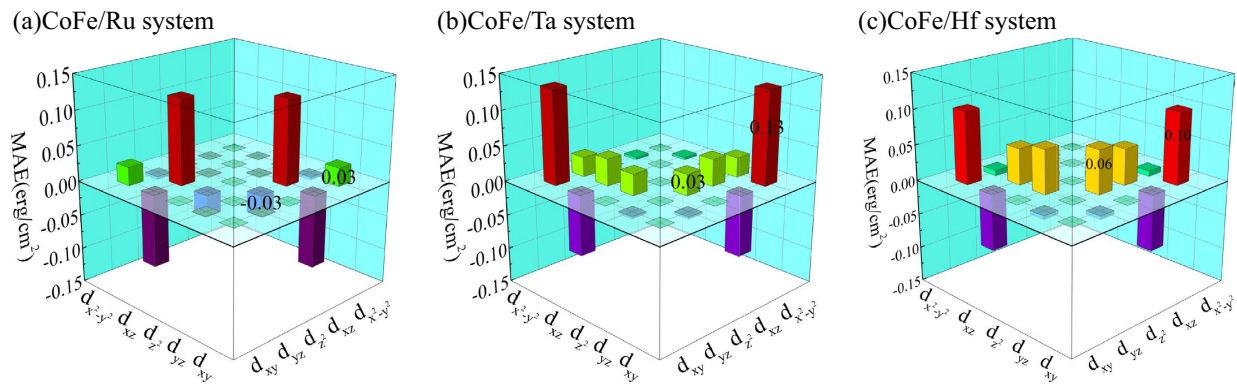

**Figure 4.** Orbital-resolved MAE of interfacial Co atoms in CoFe/X (X = Ru, Ta and Hf) systems.

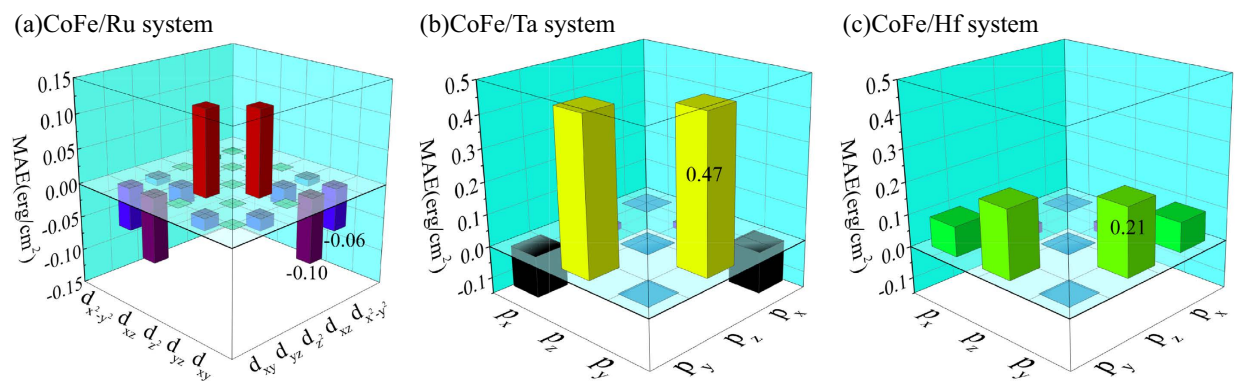

**Figure 5.** Orbital-resolved MAE of interfacial X atoms in CoFe/X (X = Ru, Ta and Hf) systems.

Generally, magnetic anisotropy can be separated into three parts, which are bulk anisotropy, interfacial anisotropy and demagnetization field[31]. Nevertheless, demagnetization field is not part of a DFT calculation. Though a large bulk MAE is predicted in the tetragonally distorted FeCo alloy[32,33], the bulk anisotropies in the MgO/CoFeB/capping layer systems are proved to be negligible in several works[8,19]. This may be due to the fact that the films are prepared by sputtering followed by an annealing, which prevents the pseudomorphic epitaxial growth of the CoFeB and removes the strain effect from the adjacent layers. In view of this, we constrain the in-plane lattice constants to that of bulk CoFe (2.83 Å) in all our calculations to remove the anisotropy induced by the strain of the CoFe layer. Take all factors above into consideration, the MAE in our work is a pure result of the interfacial anisotropy.

## Acknowledgements

W.S.Z. thanks the support by the International Collaboration Project 2015DFE12880 from the Ministry of Science and Technology in China, the support by the National Natural Science Foundation of China (Grant No. 61471015 and No.61571023) and the Beijing Municipal of Science and Technology (Grant No. D15110300320000). Q.F.Z. thanks the support by the Specialized Research Fund for the Doctoral Program of Higher Education of China (Grant No. 20131102120001) and the program for New Century Excellent Talents in University (Grant No. NCET-12-0033).

## Author Contributions

W.S.Z. coordinated the project. S.Z.P. carried out the simulations supervised by Q.F.Z. and Y.G.Z. S.Z.P., H.X.Y., L.Z., A.H., M.C. and K.L.W. performed the theoretical analysis. S.Z.P., M.X.W., J.N., J.Q.Z. and W.S.Z. wrote the manuscript. All authors interpreted the data and contributed to preparing the manuscript. Correspondence and requests for materials should be addressed to W.S.Z. and Q.F.Z.

## Additional Information

**Supplementary information** accompanies this paper at http://www.nature.com/srep

**Competing financial interests:** The authors declare no competing financial interests.

**How to cite this article**: Peng, S. *et al.* Origin of interfacial perpendicular magnetic anisotropy in MgO/CoFe/metallic capping layer structures. *Sci. Rep.* **5**, 18173; doi: 10.1038/srep18173 (2015).